\documentstyle[prl,aps]{revtex}
\begin{document}
\draft\onecolumn
\baselineskip=12pt
\title{Macroscopic Coherence for a Trapped Electron}
\author{Stefano Mancini and Paolo Tombesi}
\address{Dipartimento di Matematica e Fisica, 
Universit\`a di 
Camerino, I-62032 Camerino, Italy\\
Istituto Nazionale di Fisica Nucleare, 
Sezione di Perugia, Italy\\
and Istituto Nazionale di Fisica della Materia, 
Unit\`a di Camerino, Italy}
\date{Received: \today}
\maketitle\widetext

\begin{abstract}
We investigate the possibility of generating quantum 
macroscopic coherence phenomena by means of
relativistic effects on a trapped electron.
\end{abstract}


\narrowtext

One of the fundamental aspects of quantum mechanics is 
the exsistence of interfertence among
quantum states which signs the difference between a 
superposition of states and a mixture of states.
However, as soon as the superposition principle is 
extended to the macroscopic world the
Schr\"odinger's cat paradoxe \cite{sch} arises. 
Practically the visibility of such curious
superposition states at macroscopic level is 
precluded by the decoherence phenomena \cite{wz}.
In fact the latter results proportional to the 
"macroscopic separation" between the
superposed states \cite{zurek}.
This subject has always attracted the attention 
of physicists, but recentely, due to the improved
technology, there was a growing interest on the 
possibility of observing such superposition states,
named Schr\"odinger's cat states.

Several proposals for the generation of linear 
superpositions of 
coherent states in various nonlinear processes 
\cite{nonlinearcats} and in quantum nondemolition (QND)
measurements \cite{QNDcats} have been made.
Actually, only "mesoscopic" cats were observed in 
trapped ions and in cavity QED \cite{mesocats}.

In order to generate the cat's state in nonlinear 
systems, the crucial point is the ratio between
the strength of nonlinearity and the decoherence rate. 
The interference effects could be
preserved by slowing down the decoherence, but the 
proposed methods \cite{sqbath,feed} encounter
some difficulties in practical realization.
Hence the search for systems where such a ratio is 
sufficiently high. 

A high nonlinearity with respect to the damping, and 
consequently the dilatation of the decoherence
time, could be obtained in charged trapped systems, 
even though at microwave level. In this
letter we shall present as a system for the generation 
of cat states, an electron trapped in a
Penning trap
\cite{Penning} whose relativistic motion induces 
nonlinear effects. 
Although the relativistic correction results very 
small, its effect is however observable
\cite{releff}. The macroscopic character
relies in this case on the possibility of high 
coherent excitations of one mode of the electron
motion. We shall also suggest appropriate measurement 
techniques useful for revealing the quantum
macroscopic coherence.

In a Penning trap
one considers the motion of an electron
in a uniform magnetic field $B$ along the positive $z$ axis, 
driven by an electric field circularly polarized on the 
$x\,y$ plane and a 
static quadrupolar potential. As it is well known 
\cite{BrownG}, 
the motions of that electron in the trap are well 
separated 
in energy scale and, in a typical experimental 
situation \cite{Kells},
the interesting frequencies are 160 GHz for the 
cyclotron motion, 64 MHz 
for the axial motion, and
12 kHz for the magnetron motion. 
In what follows we shall consider only the cyclotron 
and the axial degrees of freedom neglecting the slow 
magnetron motion. 
To simplify our presentation, we assume the a priori 
knowledge of the
electron's spin, then we neglect all the
spin-related  terms in the Hamiltonian which, for an 
electron of rest mass 
$m$ and charge $-|e|$, can be approximated by 
\cite{BrownG,LT}
\begin{equation}\label{H1}
\hat H=\frac{1}{2m}\left[{\hat{\bf p}}-\frac{e}{c}
{\hat{\bf A}}\right]^2
-\frac{1}{8m^3c^2}\left[{\hat{\bf p}}-\frac{e}{c}
{\hat{\bf A}}\right]^4
+eV_0\,\frac{{\hat x}^2+{\hat y}^2-2{\hat z}^2}{4d^2}\,,
\end{equation}
with 
\begin{equation}\label{Ain}
{\hat{\bf A}}=\left(i\frac{c}{\omega_p}
\left(\epsilon^*e^{i\omega_pt}-\epsilon 
e^{-i\omega_pt}\right)-\frac{B}{2}{\hat y}\,,\,
\frac{c}{\omega_p}
\left(\epsilon^*e^{i\omega_pt}+\epsilon 
e^{-i\omega_pt}\right)+\frac{B}{2}{\hat x}\,,\,
0\right)\,,
\end{equation}
where $c$ is the 
speed of light and $\epsilon$ the amplitude of 
the driving field at angular frequency $\omega_p$;
$d$ characterizes the dimensions of the trap and 
$V_0$ is the potential applied to its electrodes. 
The second term on the r.h.s. of Eq. (\ref{H1}) 
represents the correction due to the relativistic
shift of the electron mass (we have neglected all 
contributions of higher order).
It is now convenient to introduce  the raising and 
lowering operators for the cyclotron motion,
\begin{eqnarray}\label{rlcyc}
{\hat a}&=&\frac{1}{2}\left[\beta({\hat x}-i{\hat y})+
\frac{1}{\beta\hbar}({\hat p}_y+i{\hat p}_x)\right]\,,\\
{\hat a}^{\dag}&=&\frac{1}{2}\left[
\beta({\hat x}+i{\hat y})+
\frac{1}{\beta\hbar}({\hat p}_y-i{\hat p}_x)\right]\,,
\end{eqnarray}
with $\beta=(m\omega_c/2\hbar)^{1/2}$ and 
$\omega_c=|e|B/mc$ being the cyclotron angular frequency.
Analogously, for the axial motion we define
\begin{eqnarray}\label{rlax}
{\hat a}_z&=&\left[\frac{m\omega_z}{2\hbar}\right]^{1/2}
{\hat z}+
i\left[\frac{1}{2m\hbar\omega_z}\right]^{1/2}
{\hat p}_z\,,\\
{\hat a}_z^{\dag}&=&\left[\frac{m\omega_z}{2\hbar}\right]^{1/2}
{\hat z}-
i\left[\frac{1}{2m\hbar\omega_z}\right]^{1/2}{\hat p}_z\,,
\end{eqnarray}
with $\omega_z^2=|e|V_0/md^2$. 

Thus, by using these new operators, in the dipole and rotating 
wave approximation, the
Hamiltonian (\ref{H1}) becomes
\begin{equation}\label{H2}
{\hat H}=\hbar{\hat\omega_M}({\hat a}^{\dag}{\hat a}
+\frac{1}{2})
-\hbar\mu({\hat a}^{\dag}{\hat a})^2
+i\hbar k\left(\epsilon{\hat a}^{\dag}e^{-i\omega_pt}
-\epsilon^*{\hat a}e^{i\omega_pt}\right)+
\hbar\omega_z({\hat a}^{\dag}_z{\hat a}_z+\frac{1}{2})\,,
\end{equation}
where
\begin{equation}
\mu=\frac{\hbar\omega_c^2}{2mc^2}\,,\quad k=
\frac{|e|}{\omega_p}\left(\frac{\omega_c}{2\hbar
m}\right)^{1/2}\,,
\end{equation}
and ${\hat\omega}_M$ is an operator defined by 
\begin{equation}\label{oM}
{\hat\omega_M}=\omega_c\left[1-\frac{{\hat p}_z^2}{2m^2c^2}
-\frac{\hbar\omega_c}{2mc^2}\right]\,.
\end{equation}
All terms not containing the raising or lowering operators 
have been omitted.
We have also neglected the anharmonicity of the axial motion, 
which is smaller by a factor
$(\omega_z/\omega_c)^2$ with respect to that of the 
cyclotron motion.
In the expression for ${\hat\omega_M}$ we have neglected a 
correction to the bare cyclotron angular
frequency due to the presence of the quadrupolar potential 
which is of the order 
$(\omega_z/\omega_c)^2$.
Thus, we can see in Eqs. (\ref{H2}) and (\ref{oM}) that the 
relativistic correction results in a
coupling between axial and cyclotron motions 
allowing QND measurements of cyclotron excitations 
in absence of the
pumping field \cite{Irene}, and in a further 
anharmonicity term. 

The initial state for the
cyclotron motion should be the ground state, considering 
for simplicity no excitations present in
the cyclotron motion due to the thermal bath; which however 
could be easily introduced.
Then we suppose the driving field acting initially
as a kick, and strong enough so that within its duration, 
say $\tau$, the remaining evolution can be
neglected. That could be realized if $\tau$ is
shorter than the characteristic periods, i.e. 
$\tau<<2\pi/\omega_z<<2\pi/\mu$ once one has  
chosen $\omega_p$ close enough to $\omega_c$.
Hence, we may write the effective initial cyclotron state as
\begin{equation}\label{rho0}
\hat\rho(0)={\hat D}(\alpha_0)|0\rangle\langle 0|
{\hat D}^{\dag}(\alpha_0)\,,
\end{equation}
with the displacement operator, in a frame rotating at the 
frequency $\omega_p$, given by
\begin{equation}\label{D}
\hat D(\alpha_0)=\exp\left[\alpha_0{\hat a}^{\dag}-
\alpha_0^*{\hat a}\right]\,;\quad
\alpha_0=k\epsilon\tau\,.
\end{equation}
After that the Hamiltonian governing the electron's motion 
(again in the rotating frame) is
\begin{equation}\label{H3}
{\hat H}=\hbar({\hat\omega_M}-\omega_p)
({\hat a}^{\dag}{\hat a}+\frac{1}{2})
-\hbar\mu({\hat a}^{\dag}{\hat a})^2
+\hbar\omega_z({\hat a}^{\dag}_z{\hat a}_z+\frac{1}{2})\,.
\end{equation}
Furthermore, the axial motion relaxes much faster than the 
cyclotron one \cite{BrownG},
then we can first average over the axial degrees of freedom 
and the Hamiltonian (\ref{H3}) 
simply reduces to
\begin{equation}\label{H4}
{\hat H}=\hbar(\omega_M-\omega_p)({\hat a}^{\dag}{\hat a}
+\frac{1}{2})
-\hbar\mu({\hat a}^{\dag}{\hat a})^2\,,
\end{equation}
where $\omega_M$ is no longer an operator, and is determined by 
the equilibrium temperature $T$
\begin{equation}\label{oMT}
\omega_M=\omega_c\left[1-\frac{k_BT}{2mc^2}
-\frac{\hbar\omega_c}{2mc^2}\right]\,,
\end{equation}
while the initial state of the cyclotron motion is the 
coherent state (\ref{rho0}).
The Hamiltonian (\ref{H4}) represents the same model 
studied in Ref. \cite{YS}.
By choosing $\omega_p=\omega_M$ and regarding the 
anharmonicity as the interaction part,
the discussed initial coherent state $|\alpha_0\rangle$ 
will evolve, after a time $t=\pi/2\mu$, in a
superposition of coherent states 
\begin{equation}\label{cat}
\frac{1}{\sqrt{2}}\left[
e^{-i\frac{\pi}{4}}|\alpha_0\rangle-e^{i\frac{\pi}{4}}|
-\alpha_0\rangle\right]
\end{equation}
which could be macroscopically distinguishable due to the 
opposite phase whenever a
sufficiently strong driven field $\epsilon$ is used.

However, a complete treatment of the problem has to include 
the interaction of the cyclotron motion
with the environment. The master equation for the (reduced) 
density matrix $\hat\rho$ of the
cyclotron motion can be derived by standard procedures 
\cite{qnoise} to get, in the interaction
picture,
\begin{equation}\label{mastereq}
\frac{\partial\hat\rho}{\partial t}=
i\mu\left[({\hat a}^{\dag}{\hat a})^2,\hat\rho\right]
+\frac{\gamma}{2}\left[
2\hat a\hat\rho{\hat a}^{\dag}-{\hat a}^{\dag}\hat a\hat\rho
-\hat\rho{\hat a}^{\dag}\hat a\right]\,,
\end{equation}
with $\gamma$ representing the energy relaxation rate of the 
cyclotron motion and for simplicity
the temperature of the bath is considered to be zero because 
the number of thermal excitations at
angular frequency $\omega_c$ is negligible at the usual 
temperature of performed experiments; i.e.
$T=4\,K$. Eq. (\ref{mastereq}) may be converted into a 
partial differential equation for the Husimi
\cite{Husimi} function 
$Q(\alpha,t)=\langle\alpha|\hat\rho(t)|\alpha\rangle$, 
which in turn should be solved subject to
the initial condition $Q(\alpha,0)=\exp(-|\alpha-\alpha_0|^2)$.
The solution can be found as in Ref. \cite{dami} and reads 
\begin{equation}\label{Q}
Q(\alpha,t)=e^{-|\alpha|^2-|\alpha_0|^2}
\sum_{p,q=0}^{\infty}\frac{(\alpha\alpha_0^*)^p}{p!}
\frac{(\alpha^*\alpha_0)^q}{q!}Z_{p,q}(t)\,,
\end{equation}
\begin{equation}\label{Z}
Z_{p,q}(t)=\exp\left\{
-\frac{p+q}{2}[\gamma+2i\mu(p-q)]t
+\gamma|\alpha_0|^2\frac{
1-e^{-[\gamma+2i\mu(p-q)]t}}{\gamma+2i\mu(p-q)}\right\}\,.
\end{equation}

It is worth noting that
the best way of observing the microscopic system from the outside 
world is through the measurement of the current due to the induced 
charge on the cap electrodes of the trap, as a consequence of the 
axial motion of the electron along the symmetry axis \cite{BrownG}.
Even though the relativistic effect assures a coupling between the 
cyclotron and the axial motions,
the detection models we we have in mind are based on specific 
couplings induced by Hamiltonians
immediately before the measurement process. These Hamiltonians 
could be obtained by suitable
modifications of the external fields and are
extensively discussed in Ref.\cite{PRL} where the reconstruction
of the whole Wigner function of the cyclotron state
is also considered.

Usually cat states are very fragile with respect to the 
introduction of dissipation effects,
however, the present system has the great advantage of 
obtaining a high ratio between the
nonlinearity $\mu$ and the damping coefficient $\gamma$. 
In fact the energy loss of the cyclotron
motion can be reduced by cavity effects
\cite{purcell} or by an off resonant situation \cite{GD} 
to obtain $\gamma\approx 1$ ${\rm s}^{-1}$,
so that \cite{BrownG}
$\mu/\gamma\approx 10^2>>1$.
Hence, the cat state may survive many cycles due to the
long decoherence time which is of the order of 
$(\gamma|\alpha_0|^2)^{-1}$,
as can be extracted from Eq. (\ref{Z}).

\end{document}